\documentclass{statsoc}

\usepackage[letterpaper,margin=1.25in]{geometry}
\usepackage{graphicx}
\usepackage[textwidth=10em,textsize=small]{todonotes}
\usepackage{amsmath, url}
\usepackage{natbib, color, setspace}

\title[Respondent-driven sampling]{Respondent-driven sampling: An overview in the context of human trafficking}
\author[Kunke {\it et al.}]{Jessica P. Kunke}
\address{University of Washington}
\email{jkunke@uw.edu\\}

\author{Adam Visokay}
\address{University of Washington}
\email{avisokay@uw.edu\\}

\author[Kunke {\it et al.}]{Tyler H. McCormick}
\address{University of Washington}
\email{tylermc@uw.edu\\}

\begin{document}
\begin{abstract}
Respondent-driven sampling (RDS) is both a sampling strategy and an estimation method. It is commonly used to study individuals that are difficult to access with standard sampling techniques. As with any sampling strategy, RDS has advantages and challenges.  This article examines recent work using RDS in the context of human trafficking.  We begin with an overview of the RDS process and methodology, then discuss RDS in the particular context of trafficking.  We end with a description of recent work and potential future directions.  
\end{abstract}

\section{Introduction}

Human trafficking is a global public health concern with widespread and long-lasting negative consequences.  Understanding trafficking and estimating the number of people being trafficked is complicated by the stigma, sensationalization, and secrecy of trafficking.  Recent estimates of the number of people being trafficked worldwide range from 12.3 million to 45.8 million people \citep{barrick2021advances}.  While prevalence estimation is just one of many research priorities in this field, better constraining the prevalence estimates is important for guiding policy decisions.  In this paper, we describe a common sampling technique for reaching this population, known as respondent-driven sampling (RDS).  We provide an overview of the methodology as well as a perspective on RDS in the context of trafficking.

Human trafficking was legally defined in 2000, internationally by the Palermo Protocol adopted by the UN \citep{PalermoProtocol2000}, and within the US also by the Trafficking Victims Protection Act \citep{TVPA2000}.  Generally, \textbf{human trafficking} is the use of force, fraud or coercion to exploit one or more people through commercial sex or forced labor, though inducing a minor (someone under the age of 18 years old) into commercial sex is considered human trafficking regardless of the presence of force, fraud or coercion.  There are many popular misconceptions about trafficking. For instance, human trafficking is often confused with people smuggling, in which people are moved consensually but illegally; by contrast, human trafficking can but does not necessarily involve movement, and it requires the use of force, fraud, or coercion \citep{rothman2017public,schroeder2022review}. While human trafficking has often been defined and approached through the lens of criminal justice, it is increasingly recognized as a complex public health issue.

A major challenge in studying human trafficking stems from differences in definitions or in the operationalization of the same definition from one study to another \citep{zhang2022progress}.  The Palermo Protocol explicitly mentions slavery and organ removal as forms of exploitation in its definition of trafficking, and some definitions include forced marriage, forced begging, or child soldiers.  Even under a single definition, what gets counted as trafficking on a case-by-case basis depends on popular conceptions of trafficking, which are shaped by racism, sexism, colonialism, and other systemic injustices. Much of the anti-trafficking movement in the United States, and early legislation such as the Mann Act which continues to be used in prosecutions today, has roots in unfounded early-twentieth-century panic about ``white slave traffic" \citep{allain2017whiteslave}. Black youth who trade sex for money or material needs are more likely than their white counterparts to be viewed as deviants or complicit agents rather than as victims \citep{showden2018youth}. Gendered and racialized ideas of innocence and purity misinform the popular narrative about who is trafficked, why, and what they need. This focus also encourages sensationalism, distracting from progress. Researchers, policymakers, law enforcement officials, service providers, and others who are central to identifying and combating human trafficking are subject to these same biases and misconceptions.


In addition to examining the complex question of what counts as trafficking, researchers have been working on developing effective, standardized statistical and sampling methods to understand the scope and nature of human trafficking. Many studies to-date have used administrative data, but for various reasons trafficking-related charges and prosecutions are thought to represent only a small, biased sample of existing trafficking cases. These case data reflect the aforementioned biases in determining which cases count as trafficking and who counts as a victim. Additionally, traffickers are often charged instead with other easier offenses to prove \citep{barrick2021advances}.

For this reason, many other studies have collected fresh data, which raises other challenges. Traditional survey methods assume both that researchers have a sampling frame or a list of people in the general population that includes the target population of interest (in our case, people who are being trafficked) and that respondents will willingly identify whether they are being trafficked.  In practice, accessing people in this population requires learning how to find them and building their trust. Individuals who are trafficked may not have autonomy of their movements, may distrust officials, and may not feel comfortable identifying themselves either because of stigma or fear of retribution.  In this setting, traditional sampling techniques often does not reach trafficked individuals.

One strategy to reach individuals excluded from standard surveys involves leveraging the social networks of individuals in the group of interest.  With these techniques, the researcher does not need to access a representative sample from the general population but, rather, interact with a sample of people who are \textit{connected} with members of the group of interest.  These network based methods, broadly, fall into two categories.  The first category does not involve interacting with members of the group directly.  Strategies such as the network scale-up method (NSUM) ask individuals how many people they know in the target population, i.e. people who are experiencing trafficking or who have within some time period . Responses from the general population are then ``scaled-up'' with assumptions about how well the average fraction of the respondents' networks are made up of trafficked individuals extrapolates to the population as a whole \citep{bernard1991-originalNSUM, bernard2010-nsum, killworthMcCarty1998a, laga2021thirty, mccormick_Oxford_chapter}.

These indirect methods have the advantage that they do not require respondents to identify themselves or specific other people as members of the group of interest.  A disadvantage is that these methods are often limited to prevalence estimation rather than gaining additional insight into risk factors, experiences, or possible paths out of trafficking. NSUM-based trafficking prevalence estimates have not yet been published to our knowledge, but several studies as part of the Prevalence Reduction Innovation Forum (PRIF) initiative are currently underway to directly compare NSUM and other estimation methods on the same target populations \citep{PRIF2023, schroeder2022review, zhang2022progress}.

A second class of methods, which we focus on in this paper, involve directly interacting with individuals in the group of interest.  These methods fall under a general class of methods known as ``link-tracing'' or ``chain-referral'' designs because recruitment proceeds along links in the social network connecting individuals who are victims of trafficking.  Several related iterations of link-tracing designs have been proposed, so we begin with some terminology. {\bf Chain-referral sampling} is a general term for a method that ``traces'' respondents' networks as a means of recruitment. {\bf Snowball sampling} is a chain-referral method originally proposed as a way to learn about network features that starts with a probability sample of respondents and traces their networks. {\bf Respondent-driven sampling (RDS)} is sometimes called ``non-probability snowball sampling'' because it starts with a \emph{convenience sample} and respondents choose network members to recruit.  Researchers use RDS to estimate prevalence, understand characteristics of particular population (e.g. the fraction of sex workers in an area who have been trafficked), or to access members of a hard-to-reach group for an intervention.  

This paper focuses on RDS since it is an increasingly popular sampling and estimation strategy for human trafficking.  In their scoping review of measurement strategies to learn about the prevalence and experience of human trafficking, \cite{barrick2021advances} report that around 16\% of the studies included in their review utilized RDS. \cite{franchino2022prevalence} conducted a scoping review on prevalence estimates for domestic minor sex trafficking and commercial sexual exploitation of children, and of the six studies included in the review, one used RDS. These examples demonstrate both that RDS is being used in human trafficking research and that the methodology is not yet standardized.

In the remainder of this paper, we provide an overview of the implementation and assumptions required for RDS (Section~\ref{sec:rds}), followed by a discussion of RDS methodology (Section~\ref{sec:rdsmethod}).  We then turn to discussion of RDS and its potential advantages and challenges in the context of human trafficking (Section~\ref{sec:rdstraf}), as well as recent advances and potential opportunities (Section~\ref{sec:next}). Finally, we conclude in Section~\ref{sec:concl}.

\section{RDS implementation and assumptions}
\label{sec:rds}
In this section, we describe the typical RDS process and assumptions required for statistical inference, which we discuss in the next section.  RDS can be used for three distinct purposes: (i) estimating the characteristics of the group of individuals (e.g. the fraction of trafficking victims who are also minors); (ii) estimating a population size or prevalence; and (iii) accessing a representative sample of individuals for further study or intervention (e.g. to evaluate the effectiveness a particular type of outreach).  RDS relies on members of the group of interest to recruit other members of the group into the study, thus leveraging the social connections and relationships of group members to increase participation.






\begin{figure}
    \centering
    \includegraphics[width=\textwidth]{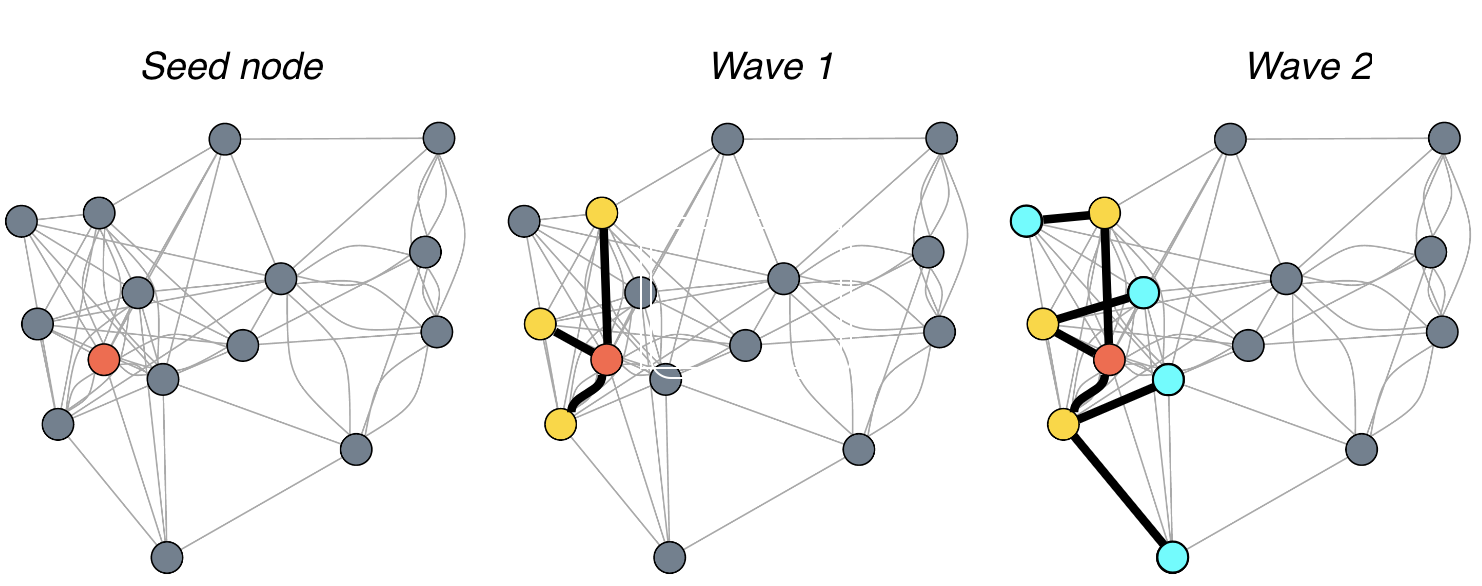}
    \caption{Each figure shows an entire example network. Nodes recruited in waves 1, 2 and 3 are shown in red, yellow, and light blue, respectively. Grey nodes are never recruited, and bolded paths indicate directed recruitment links. Paths that are not bolded remain unobserved to the researcher.}
    \label{fig:rdsexample}
\end{figure}

The RDS process begins with a convenience sample of members of the group of interest, people who have been trafficked within some period of time.  These individuals are often known to researchers from previous studies or interventions, or they have previously interacted with public health infrastructure.  These initial recruits are known as \textbf{seeds}.  The seed individuals are asked to recruit a particular number of additional individuals, also from the group of interest.  Each of these recruits typically receives an incentive to participate in the study, known as the \textbf{primary incentive}.  Recruits are then asked to bring more individuals from the group of interest into the study.  When a new recruit participates in the study, the person who recruited them receives a \textbf{secondary incentive}. Each new recruitment cycle defines a \textbf{wave} of a recruitment \textbf{chain}, the complete set of individuals and their referral connections including the original seeds.

There are a number of practical considerations when performing RDS.  The order of recruitment can be consequential when analyzing RDS, as we discuss below, and recording it is necessary if there is a secondary incentive.  Thus, it is critical for researchers to keep track of the recruitment sequencing.  Often, this task is accomplished by passing out coupons with unique identifying numbers.  Each participant gets a certain number of coupons containing their unique recruiter number, so that when a new participant brings a coupon they were given, the study team knows who recruited them.  An additional consideration is the number of coupons available to each person.  Allowing each person to recruit more people reduces the chances that a chain terminates in the early waves.  For a given sample size, however, it also means that the study can incorporate fewer seeds and, thus, have chains originating in potentially different parts of the network.

Figure~\ref{fig:rdsexample} demonstrates the RDS process on a small example network, starting with a single seed (RDS typically begins with multiple seeds). We begin with a single seed, selected as a convenience sample, denoted in red.  The seed recruits, in this example, three additional participants in the first wave.  These participants then, in turn, recruit additional participants in subsequent waves. The unbolded network edges in the figure are not observed. This example also illustrates two challenges with the RDS procedure.  First, the sampling is happening on top of an existing social network which is unknown to the researcher and difficult to recover from the RDS chain.  Second, the sampling process on that network is controlled by the respondents, not by the researcher, meaning that the choice of who is included in the study is up to the respondents and may or may not be representative or meet other desirable sample criteria.

With this procedure in mind, we begin our discussion of the statistical properties and assumptions of RDS.  Since the initial seeds for RDS are a convenience sample, they are not representative of the population.  In an ideal world, however, subsequent recruitment waves would ``move away'' from the initial seeds in the social network, making it less and less consequential which initial seeds are used.  Also under ideal circumstances, as the chains traverse the network they will include respondents with heterogeneous characteristics and the frequency of those characteristics in the sample will be roughly that of the population.  If this happens, RDS behaves like a mathematical process known as a Markov Chain Monte Carlo \citep{goel2009respondent}.

This ideal behavior of RDS requires several assumptions.  First, RDS assumes that members of the population can be reached through their network, i.e. that they know one another reciprocally, interact frequently, are willing to recruit others, and have mobility. As we discuss further later, this assumption may not be met in the context of human trafficking. Restrictions on mobility, for example, may make it impossible for individuals to receive a coupon or to bring a coupon they may receive to a research center and participate in the study.  Second, RDS assumes that respondents' network sizes are either known or accurately estimated. This assumption is necessary because the likelihood of being sampled depends on the respondent's network; a person with more contacts has more chances to be included.  Third, the sampling process needs to continue through enough waves to mitigate the dependence on seeds.  If there aren't enough waves, then the structure of the sample will be too closely related to the initial seeds. \cite{gile2010respondent} found that the seed-induced bias depends on the extent of homophily and the number of sampling waves. If there are substantial bottlenecks in the network then the recruitment process can get ``stuck'' in one pocket not explore the full extent of the graph (see \cite{rohe2019design}, for example).  Particularly with small groups, the sample size can become close to the total population size. Fourth, RDS assumes sampling is done with replacement, meaning that the study may recruit the same person more than once.  Fifth, RDS assumes that network connections are reciprocal, that person A is equally likely to refer person B as person B would be to refer person A \citep{volz2008probability}.

Finally, RDS assumes that respondents recruit randomly from their contacts. Under this assumption, the only factor that impacts how likely you are to be recruited is your number of contacts.  In practice, though, recruitment may be highly preferential, or even if the recruitment is random, there may be selection bias in which people who receive coupons are more likely to actually participate in the study. Preferential referral can introduce bias \citep{gile2010respondent}. The recruitment process is likely based on several factors that are not visible to the researchers and therefore cannot be controlled for in a straightforward way. However, in Section \ref{sec:next}, we discuss some extensions of RDS that are designed to give the researcher more control over recruitment. \cite{goel2010assessing} point out that violations of these assumptions can lead to substantial issues with statistical inference.

\section{RDS estimation}
\label{sec:rdsmethod}

As mentioned previously, researchers use RDS for a variety of estimation goals.  In this section, we discuss briefly the intuition behind using RDS to estimate prevalence and using RDS to estimate a population fraction.  We focus on estimating the population fraction.  Readers interested in prevalence estimation can refer to to~\cite{handcock2014estimating} or~\cite{crawford2018hidden}. 

We take as a working example conducting RDS to estimate the fraction of sex workers who have been trafficked.  The researcher performs RDS on the population of sex workers (which is likely difficult to access with other sampling methods due to stigma, fear of prosecution, or other factors) and for each person recruited performs an interview where the person indicates whether they have been trafficked.

The first estimator that we might consider would be to simply take the average. That is, we take the number of sex workers recruited who report being trafficked divided by the total number recruited.  This estimator would be biased because some people are more likely to be recruited than others.  Specifically, as described in the previous sections, people with more contacts have more chances to be included in the sample.  To compensate for this, a class of estimators called \emph{Horvitz-Thompson estimators} (or sometimes generalized Horvitz-Thompson estimators) re-weight the average by the inverse of the likelihood that a person is included in the sample~\citep{heckathorn1997respondent,salganik20045,volz2008probability}. Respondents with fewer connections are less likely to be included in a referral chain, and thus have a lower inclusion probability, so the estimator gives their responses extra weight, proportional to how likely (or not) they are to be referred. In this case, the RDS estimator uses the inverse of each respondent's estimated degree\textemdash how many reciprocal ties they have to other members of the population of interest\textemdash as a correction factor for the estimate. ~\cite{gile2010respondent} provide a much more thorough discussion of these estimators and their properties. The question of how to measure the uncertainty in these estimators is also an area of current work (see for example~\cite{green2020consistency, rohe2019design,treeboot2016, goel2010assessing}). 




\section{Considerations in the context of trafficking}
\label{sec:rdstraf}


There are several considerations for successfully implementing RDS, particularly in the context of human trafficking. Beginning with observations from \cite{simic2006exploring}, we illustrate some of the difficulties that have arisen in the context of specific trafficking studies from the literature. In their studies of sex workers in three different countries, \cite{simic2006exploring} were unable to recruit enough study participants through RDS, and they attributed this to several interrelated potential factors:

\textbf{Lack of trust.} General mistrust of official agencies combined with the sensitivity around sex worker status to reduce participation. Even though the study team worked in advance to build trust and create community ties, tight control by brothels and police crackdowns increased potential participants' reluctance to identify themselves or others. Some participants did not want to reveal their own status as a sex worker to others by recruiting them, since the recruits would find out the eligibility criteria of the study when they were interviewed. Participants sometimes avoided recruiting particular people to avoid identifying them.  \cite{simic2006exploring} suggest it may help to run the studies for a longer time to gain more trust, especially if the seed sampling and interviews occur at a location with ongoing services that can be used as an interview site. However, in places where sex workers have little contact with local services, this may not be feasible.

\textbf{Social network structure.} RDS assumes dense, connected networks. By contrast, many of the sex workers in these communities were isolated, either due to restricted movement or because they worked independently and did not tend to reveal their status to others. In Serbia \cite{simic2006exploring} found that most sex workers worked independently and did not connect much with each other. They also found that street sex workers, organized sex workers, and independent sex workers tended not to connect with each other. Street sex workers tended to be socially separated by ethnicity, sexuality, and other aspects of identity.

\textbf{Restricted movement.} In Montenegro, brothels were tightly controlled and sex workers were not allowed to leave. Sex workers with restricted movement were unlikely to be able to receive a coupon or to go to the study locations to participate even if they received a coupon. This seems to have combined with intense policing practices in retaliation for a recent police HIV infection to hamper study recruiting.

\textbf{Inadequate study incentives.} The financial incentives provided were not high enough relative to the earnings sex workers could make and the opportunity cost of missing work to participate in the study. Sex workers were more interested in the HIV testing than the financial incentives. If the incentive is too high and generally appealing beyond the target population, this can encourage people who are not trafficked to attempt to participate. This illustrates the importance of better identifying effective incentives in advance of the study.

We note a few additional concerns illustrated by other studies:

\textbf{Unrepresentative seeds.} \cite{zhang2014SanDiego} implemented RDS to estimate trafficking prevalence among unauthorized migrant laborers in San Diego. They worked with a community partner to build trust with the migrant community and recruit initial respondents, and they were able to exceed their estimated minimum sample size. However, the study was also limited by the combination of ethnic homophily and an ethnically homogeneous set of seeds. All their seeds were Mexican workers, and unauthorized Mexican migrants tended to connect among themselves rather than with other Spanish-speaking unauthorized migrants; as a result, the study had limited recruitment among non-Mexicans and was not representative of the undocumented migrant laborer community as a whole.

\textbf{Non-reciprocal links.} In a study of sex trafficking risk factors using data from a 2011 RDS survey of urban street-based Ohio sex workers, \cite{chohaney2016minor} found that the network violated the reciprocity assumption. Roughly one quarter of respondents described the person who referred them as a stranger, one quarter described them as neighbors or someone they ``kind of know,” and the remaining half described them as friends or family members.

\textbf{Lack of visibility to other sex workers.} While \cite{carrillo2020implementing} did not study trafficking but rather women who exchange sexual services for money or other goods, their study findings are relevant to implementing RDS in trafficking studies as well. One-third of the study participants recruited by previous waves of participants turned out to be ineligible or had not exchanged sex in the past year.  One potential reason they identified is that women may have limited knowledge of whether specific other women they know are exchanging sex.

\textbf{Misalignment between recruitment and eligibility criteria.} Additionally in the study by \cite{carrillo2020implementing}, recruiters were asked to recruit women who exchanged sex, not women who exchanged sex specifically within the last 12 months, which was the specific question that interviewers asked to assess group membership. This illustrates the importance of aligning the recruitment instructions with the study criteria, and yet asking participants to confirm more specific criteria when they recruit further participants may not be feasible if they require them to ask others more specific and sensitive questions.

\section{Recent advances and future directions}
\label{sec:next}

In this section we discuss three extensions of RDS that were developed to address the challenges that arise when some or many of the critical assumptions mentioned in Section \ref{sec:rds} are not met in practice: network sampling with memory (NSM), randomized respondent-driven sampling (RRDS), and link-tracing sampling methods that combine various approaches to improve estimation. 

\subsection{Network Sampling with Memory (NSM)}

NSM is an application of RDS that builds upon advancements in the mathematics and computer science literature on random walks on graphs proposed by \cite{mouw2012NSM}. At a high level, the researcher supervises and strategically directs the recruitment process as it unfolds. This gives the researcher more control, ultimately yielding a more efficient sampling framework than can be attained with traditional RDS. 

NSM begins with initial seeds from a convenience sample, all of whom provide a roster of their contacts known to be members of the target population in addition to answering the substantive interview questions chosen by the researcher. NSM is then implemented as a two-step approach\textemdash the \textbf{Search} mode followed by the \textbf{List} mode. Search mode prioritizes \textbf{bridge} nodes, which are individuals who connect two or more clusters together, to sufficiently explore the network, while List mode which ensures that nodes sampled early in the process are not over-represented in the sample.

Search mode takes the network information of respondents and uses the local topography to identify bridge nodes that connect unexplored portions of the network. These nodes are then given priority in the recruitment process. The researcher pre-specifies a threshold that triggers when the network has been sufficiently explored by Search mode. After Search mode concludes, NSM proceeds to the List mode which entails two steps: (1) keeping a list of all individuals on the revealed network and (2) sampling from that list with the same cumulative probability for each individual such that new additions to the list are given priority. 

One of the key advantages to NSM compared to RDS is the improved efficiency in searching the network. Given high quality network data collected from each respondent, the computation and processing costs associated with this method are small. However, collecting high quality network data from human populations can be prohibitively expensive or logistically infeasible. NSM may also be impractical in the context of human trafficking  where estimates of each respondents network degree are be highly variable. There are also additional time and effort costs associated with the real-time supervision and direction of the recruitment process that make NSM more challenging to implement in practice.

\subsection{Randomized Respondent-Driven Sampling (RRDS)}

Respondents recruiting randomly from their contacts is an important assumption of RDS that is often violated in practice. 

\begin{figure}
    \centering
    \includegraphics[width=\textwidth]{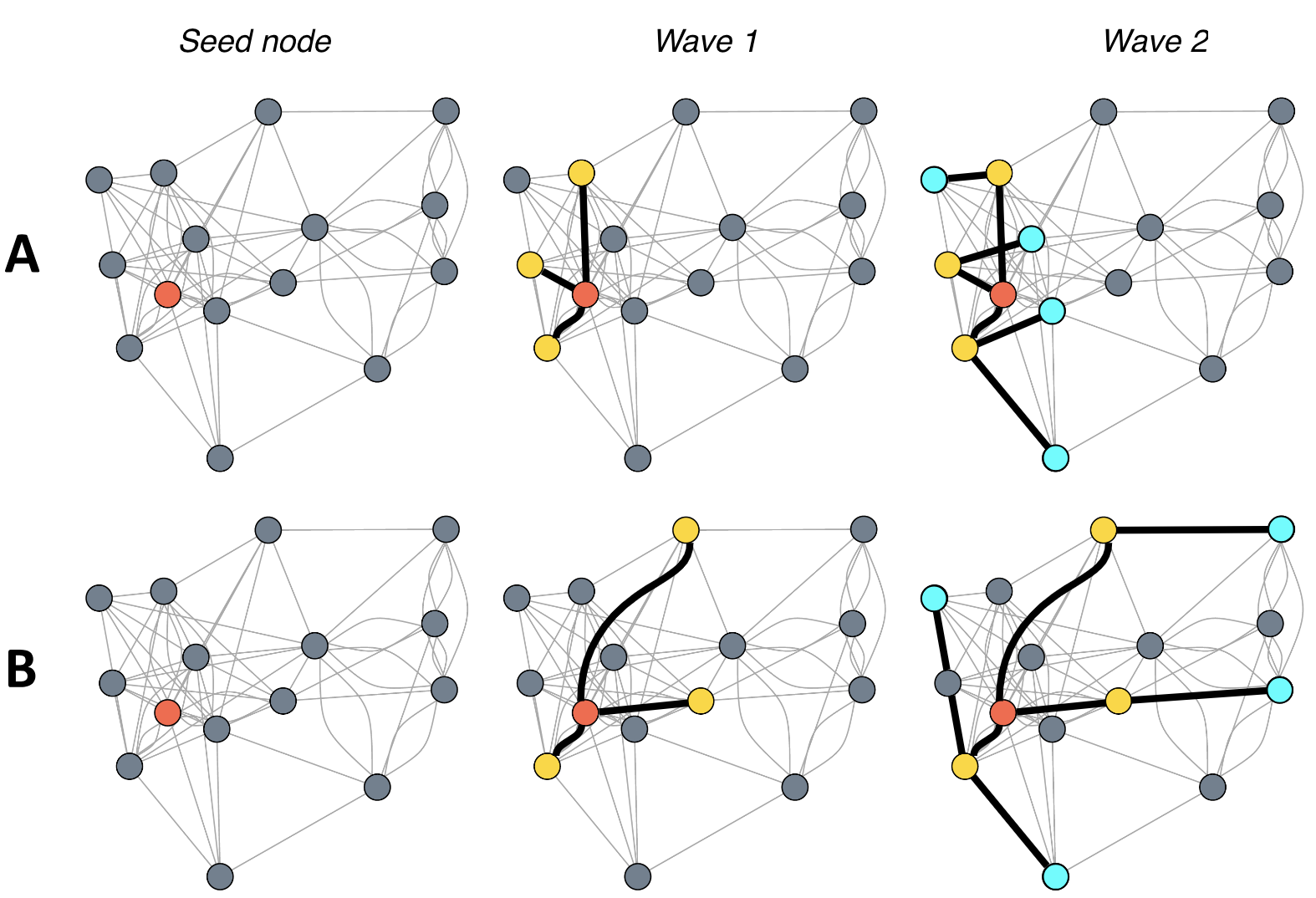}
    \caption{Each figure shows an entire example network. Nodes recruited in waves 1, 2 and 3 are shown in red, yellow, and light blue, respectively. Grey nodes are never recruited, and bolded paths indicate directed recruitment links. (A) Demonstrates the process of a traditional RDS recruitment tree. (B) Shows how randomizing recruitment of respondents' contacts for each wave results in better coverage.}
    \label{fig:rrdsexample}
\end{figure}

~\cite{rrds} propose a cell-phone based variant of RDS that addresses this challenge.  
The set up is similar to RDS in that researchers begin with a convenience sample for the initial seeds. From each seed, the researcher collects a list of phone numbers for their contacts believed to also be in the target population. The researcher then chooses a random subset of the respondents' contacts, administers the survey, and collects a list of phone numbers of their contacts for the subsequent wave. This process is then repeated until the desired sample size is reached. This process is illustrated in Figure \ref{fig:rrdsexample}. 

RRDS has several advantages relative to traditional RDS.

{\bf More reliable randomization.} By introducing randomness at each wave of the recruitment process, this method is much closer to the ideal RDS assumption of random selection among contacts. 

{\bf Phone based rather than venue based.} RRDS is a carried out using phone surveys and does not require in person interview sites. This makes RRDS an attractive option for applications like human trafficking where restricted mobility makes it difficult to recruit respondents in person. 

{\bf More control over recruitment.} While RRDS is still respondent driven, introducing randomization gives the researcher more control over the recruitment process compared to traditional RDS.

{\bf Simple to supervise.} The time and effort costs associated with administering the randomization are relatively low. It is as simple as drawing a simple random sample from each respondents list of phone numbers before proceeding with the next wave of recruitment.

RRDS will, of course, not be well-suited to every context.  The most obvious requirement is the need to work with a population where individuals have access to cellphones (and the autonomy to use those phones as they wish).  Individuals in the group of interest without access to a phone will be excluded.  RRDS also relies on individuals knowing phone numbers (or saving contacts) of other members of the group of interest. ~\cite{rrds} implemented RRDS with workers in a large-scale industrial manufacturing setting during the peak of the COVID-19 pandemic.  Telephone surveys were the only means to obtain critical information on the health and well-being of workers, as in-person activities were restricted.  More research is needed to understand how effective this approach will be in the context of trafficking. 



\subsection{Combining link-tracing methods and other strategies}

Specifically in the context of trafficking studies, there have also been efforts to combine RDS with traditional representative sampling methods, venue-based sampling, and multiple systems estimation to improve estimation and inference \citep{vincent2017estimating, vincent2018recent}. \cite{zhang2019victims} leverage overlaps among RDS participants' social networks to improve population size estimation using mark-recapture methods. \cite{vincent2021kathmandu_MSE} conduct both RDS and venue-based sampling, then use mark-recapture methodology to improve inference by combining the two samples. \cite{vincent2021india_linktracing} increase the number of initial seeds and reduce the necessary number of RDS waves.


\section{Conclusion}
\label{sec:concl}
Human trafficking is a complex, stigmatized, secretive and constantly evolving phenomenon.  Respondent-driven sampling offers several advantages over traditional survey methods for studying this population, but its success requires building trust with the relevant communities, providing well-informed and motivating incentives, and understanding and accounting for several aspects of the social network structure of the people being trafficked and their surrounding community. We have highlighted some potential suggestions from the literature for addressing these concerns as well as promising directions for future research.




\bibliographystyle{rss}
\bibliography{trafficking_RDS}

\begin{thebibliography}{37}
\expandafter\ifx\csname natexlab\endcsname\relax\def\natexlab#1{#1}\fi
\expandafter\ifx\csname url\endcsname\relax
  \def\url#1{\texttt{#1}}\fi
\expandafter\ifx\csname urlprefix\endcsname\relax\def\urlprefix{URL: }\fi

\bibitem[{Allain(2017)}]{allain2017whiteslave}
Allain, J. (2017) White slave traffic in international law.
\newblock \textit{Journal of Trafficking and Human Exploitation}, \textbf{1}, 1--40.

\bibitem[{Baraff et~al.(2016)Baraff, McCormick and Raftery}]{treeboot2016}
Baraff, A.~J., McCormick, T.~H. and Raftery, A.~E. (2016) Estimating uncertainty in respondent-driven sampling using a tree bootstrap method.
\newblock \textit{Proceedings of the National Academy of Sciences}, \textbf{113}, 14668--14673.
\newblock \urlprefix\url{https://www.pnas.org/doi/abs/10.1073/pnas.1617258113}.

\bibitem[{Barrick and Pfeffer(2021)}]{barrick2021advances}
Barrick, K. and Pfeffer, R.~J. (2021) Advances in measurement: A scoping review of prior human trafficking prevalence studies and recommendations for future research.
\newblock \textit{Journal of Human Trafficking}, \textbf{8}, 193--211.

\bibitem[{Bernard et~al.(2010)Bernard, Hallett, Iovita, Johnsen, Lyerla, McCarty, Mahy, Salganik, Saliuk, Scutelniciuc, Shelley, Sirinirund, Weir and Stroup}]{bernard2010-nsum}
Bernard, H.~R., Hallett, T., Iovita, A., Johnsen, E.~C., Lyerla, R., McCarty, C., Mahy, M., Salganik, M.~J., Saliuk, T., Scutelniciuc, O., Shelley, G.~A., Sirinirund, P., Weir, S. and Stroup, D.~F. (2010) Counting hard-to-count populations: The network scale-up method for public health.
\newblock \textit{Sexually Transmitted Infections}, \textbf{86}, 1368--4973.

\bibitem[{Bernard et~al.(1991)Bernard, Johnsen, Killworth and Robinson}]{bernard1991-originalNSUM}
Bernard, H.~R., Johnsen, E.~C., Killworth, P.~D. and Robinson, S. (1991) Estimating the size of an average personal network and of an event subpopulation: Some empirical results.
\newblock \textit{Social Science Research}, \textbf{20}, 109--121.

\bibitem[{Boudreau et~al.(2023)Boudreau, Heath, McCorimck and Visokay}]{rrds}
Boudreau, L., Heath, R., McCorimck, T.~H. and Visokay, A. (2023) Randomized respondent-driven sampling: A cellphone based approach to sampling in disaster settings.
\newblock \textit{Working Paper, Univeristy of Washington}.

\bibitem[{Carrillo et~al.(2020)Carrillo, Rivera and Braunstein}]{carrillo2020implementing}
Carrillo, S.~A., Rivera, A.~V. and Braunstein, S.~L. (2020) Implementing respondent-driven sampling to recruit women who exchange sex in {New York City}: Factors associated with recruitment and lessons learned.
\newblock \textit{AIDS and Behavior}, \textbf{24}, 580--591.

\bibitem[{Chohaney(2016)}]{chohaney2016minor}
Chohaney, M.~L. (2016) Minor and adult domestic sex trafficking risk factors in {Ohio}.
\newblock \textit{Journal of the Society for Social Work and Research}, \textbf{7}, 117--141.

\bibitem[{Crawford et~al.(2018)Crawford, Wu and Heimer}]{crawford2018hidden}
Crawford, F.~W., Wu, J. and Heimer, R. (2018) Hidden population size estimation from respondent-driven sampling: a network approach.
\newblock \textit{Journal of the American Statistical Association}, \textbf{113}, 755--766.

\bibitem[{Franchino-Olsen et~al.(2022)Franchino-Olsen, Chesworth, Boyle, Rizo, Martin, Jordan, Macy and Stevens}]{franchino2022prevalence}
Franchino-Olsen, H., Chesworth, B.~R., Boyle, C., Rizo, C.~F., Martin, S.~L., Jordan, B., Macy, R.~J. and Stevens, L. (2022) The prevalence of sex trafficking of children and adolescents in the {United States}: A scoping review.
\newblock \textit{Trauma, Violence, \& Abuse}, \textbf{23}, 182--195.

\bibitem[{Gile and Handcock(2010)}]{gile2010respondent}
Gile, K.~J. and Handcock, M.~S. (2010) Respondent-driven sampling: an assessment of current methodology.
\newblock \textit{Sociological Methodology}, \textbf{40}, 285--327.

\bibitem[{Goel and Salganik(2009)}]{goel2009respondent}
Goel, S. and Salganik, M.~J. (2009) Respondent-driven sampling as {Markov chain Monte Carlo}.
\newblock \textit{Statistics in Medicine}, \textbf{28}, 2202--2229.

\bibitem[{Goel and Salganik(2010)}]{goel2010assessing}
--- (2010) Assessing respondent-driven sampling.
\newblock \textit{Proceedings of the National Academy of Sciences}, \textbf{107}, 6743--6747.

\bibitem[{Green et~al.(2020)Green, McCormick and Raftery}]{green2020consistency}
Green, A., McCormick, T. and Raftery, A. (2020) Consistency for the tree bootstrap in respondent-driven sampling.
\newblock \textit{Biometrika}, \textbf{107}, 497--504.

\bibitem[{Handcock et~al.(2014)Handcock, Gile and Mar}]{handcock2014estimating}
Handcock, M.~S., Gile, K.~J. and Mar, C.~M. (2014) Estimating hidden population size using respondent-driven sampling data.
\newblock \textit{Electronic journal of statistics}, \textbf{8}, 1491.

\bibitem[{Heckathorn(1997)}]{heckathorn1997respondent}
Heckathorn, D.~D. (1997) Respondent-driven sampling: a new approach to the study of hidden populations.
\newblock \textit{Social problems}, \textbf{44}, 174--199.

\bibitem[{Killworth et~al.(1998)Killworth, Johnsen, McCarty, Shelley and Bernard}]{killworthMcCarty1998a}
Killworth, P.~D., Johnsen, E.~C., McCarty, C., Shelley, G.~A. and Bernard, H.~R. (1998) A social network approach to estimating seroprevalence in the {U}nited {S}tates.
\newblock \textit{Social Networks}, \textbf{20}, 23--50.

\bibitem[{Laga et~al.(2021)Laga, Bao and Niu}]{laga2021thirty}
Laga, I., Bao, L. and Niu, X. (2021) Thirty years of the network scale-up method.
\newblock \textit{Journal of the American Statistical Association}, \textbf{116}, 1548--1559.

\bibitem[{McCormick(2021)}]{mccormick_Oxford_chapter}
McCormick, T.~H. (2021) The network scale-up method.
\newblock In \textit{{The Oxford Handbook of Social Networks}}. Oxford University Press.
\newblock \urlprefix\url{https://doi.org/10.1093/oxfordhb/9780190251765.013.14}.

\bibitem[{Mouw and Verdery(2012)}]{mouw2012NSM}
Mouw, T. and Verdery, A.~M. (2012) Network sampling with memory: A proposal for more efficient sampling from social networks.
\newblock \textit{Sociological Methodology}, \textbf{42,}, 206--256.

\bibitem[{Rohe(2019)}]{rohe2019design}
Rohe, K. (2019) {A critical threshold for design effects in network sampling}.
\newblock \textit{The Annals of Statistics}, \textbf{47}, 556 -- 582.
\newblock \urlprefix\url{https://doi.org/10.1214/18-AOS1700}.

\bibitem[{Rothman et~al.(2017)Rothman, Stoklosa, Baldwin, Chisolm-Straker, Kato~Price, Atkinson and Trafficking}]{rothman2017public}
Rothman, E.~F., Stoklosa, H., Baldwin, S.~B., Chisolm-Straker, M., Kato~Price, R., Atkinson, H.~G. and Trafficking, H. (2017) Public health research priorities to address {US} human trafficking.

\bibitem[{Salganik and Heckathorn(2004)}]{salganik20045}
Salganik, M.~J. and Heckathorn, D.~D. (2004) 5. sampling and estimation in hidden populations using respondent-driven sampling.
\newblock \textit{Sociological methodology}, \textbf{34}, 193--240.

\bibitem[{Schroeder et~al.(2022)Schroeder, Edgemon, Aletraris, Kagotho, Clay-Warner and Okech}]{schroeder2022review}
Schroeder, E., Edgemon, T.~G., Aletraris, L., Kagotho, N., Clay-Warner, J. and Okech, D. (2022) A review of prevalence estimation methods for human trafficking populations.
\newblock \textit{Public Health Reports}, \textbf{137}, 46S--52S.

\bibitem[{Showden and Majic(2018)}]{showden2018youth}
Showden, C.~R. and Majic, S. (2018) \textit{Youth who trade sex in the US: Intersectionality, agency, and vulnerability}.
\newblock Temple University Press.

\bibitem[{Simic et~al.(2006)Simic, Johnston, Platt, Baros, Andjelkovic, Novotny and Rhodes}]{simic2006exploring}
Simic, M., Johnston, L.~G., Platt, L., Baros, S., Andjelkovic, V., Novotny, T. and Rhodes, T. (2006) Exploring barriers to `respondent driven sampling' in sex worker and drug-injecting sex worker populations in {Eastern Europe}.
\newblock \textit{Journal of Urban Health}, \textbf{83}, 6--15.

\bibitem[{{United Nations General Assembly}(2000)}]{PalermoProtocol2000}
{United Nations General Assembly} (2000) Protocol to prevent, suppress and punish trafficking in persons especially women and children, supplementing the {United Nations Convention against Transnational Organized Crime}.
\newblock \urlprefix\url{https://www.ohchr.org/en/instruments-mechanisms/instruments/protocol-prevent-suppress-and-punish-trafficking-persons}.

\bibitem[{{University of Georgia Center on Human Trafficking Research and Outreach}(2023)}]{PRIF2023}
{University of Georgia Center on Human Trafficking Research and Outreach} (2023) {Prevalence Reduction Innovation Forum}.
\newblock \urlprefix\url{https://cenhtro.uga.edu/prif/about_prif/}.

\bibitem[{{Victims of Trafficking and Violence Protection Act}(2000)}]{TVPA2000}
{Victims of Trafficking and Violence Protection Act} (2000) {Pub. L. No. 106-386}.
\newblock \urlprefix\url{https://www.govinfo.gov/app/details/PLAW-106publ386}.

\bibitem[{Vincent(2018)}]{vincent2018recent}
Vincent, K. (2018) Recent advances on estimating population size with link-tracing sampling.
\newblock \textit{arXiv preprint arXiv:1709.07556}.

\bibitem[{Vincent et~al.(2021{\natexlab{a}})Vincent, Dank, Jackson, Zhang and Liu}]{vincent2021kathmandu_MSE}
Vincent, K., Dank, M., Jackson, O., Zhang, S.~X. and Liu, W. (2021{\natexlab{a}}) Estimating young women working in {K}athmandu’s adult entertainment sector: A hybrid application of respondent driven sampling and venue site sampling.
\newblock \textit{Journal of Human Trafficking}, 1--16.

\bibitem[{Vincent and Thompson(2017)}]{vincent2017estimating}
Vincent, K. and Thompson, S. (2017) Estimating population size with link-tracing sampling.
\newblock \textit{Journal of the American Statistical Association}, \textbf{112}, 1286--1295.

\bibitem[{Vincent et~al.(2021{\natexlab{b}})Vincent, Zhang and Dank}]{vincent2021india_linktracing}
Vincent, K., Zhang, S.~X. and Dank, M. (2021{\natexlab{b}}) Searching for sex trafficking victims: Using a novel link-tracing method among commercial sex workers in {Muzaffarpur, India}.
\newblock \textit{Crime \& Delinquency}, \textbf{67}, 2254--2277.

\bibitem[{Volz and Heckathorn(2008)}]{volz2008probability}
Volz, E. and Heckathorn, D.~D. (2008) Probability based estimation theory for respondent driven sampling.
\newblock \textit{Journal of official statistics}, \textbf{24}, 79.

\bibitem[{Zhang(2022)}]{zhang2022progress}
Zhang, S.~X. (2022) Progress and challenges in human trafficking research: Two decades after the {Palermo} protocol.
\newblock \textit{Journal of Human Trafficking}, \textbf{8}, 4--12.

\bibitem[{Zhang et~al.(2019)Zhang, Dank, Vincent, Narayanan, Bharadwaj and Balasubramaniam}]{zhang2019victims}
Zhang, S.~X., Dank, M., Vincent, K., Narayanan, P., Bharadwaj, S. and Balasubramaniam, S.~M. (2019) Victims without a voice: Measuring worst forms of child labor in the {Indian State of Bihar}.
\newblock \textit{Victims \& Offenders}, \textbf{14}, 832--858.

\bibitem[{Zhang et~al.(2014)Zhang, Spiller, Finch and Qin}]{zhang2014SanDiego}
Zhang, S.~X., Spiller, M.~W., Finch, B.~K. and Qin, Y. (2014) Estimating labor trafficking among unauthorized migrant workers in {San Diego}.
\newblock \textit{The Annals of the American Academy of Political and Social Science}, \textbf{653}, 65--86.

\end{thebibliography}
\end{document}